\documentclass[conference, a4paper]{IEEEtran}

\usepackage{pgfplots}
  \pgfplotsset{compat=newest}

\IEEEoverridecommandlockouts
\usepackage{main}
\begin{document}
\begin{acronym}
    \acro{AP}{access point}
    \acro{PTP}{Precision Time Protocol}
    \acro{OWD}{One-way delay}
    \acro{RTT}{Round-trip time}
    \acro{SMS}{Smart manufacturing systems}
    \acro{CPS}{Cyber-physical systems}
    \acro{IIoT}{Industrial Internet-of-Things}
    \acro{TUB}{Time Uncertainty Bound}
    \acro{UTC}{Coordinated Universal Time}
    \acro{GPS}{Global Positioning System}
    \acro{OWD}{One-way delay}
    \acro{ppm}{parts per million}
    \acro{RTT}{Round trip time}
    \acro{NIC}{Network interface controllers}
    \acro{DES}{Discrete event simulator}
    \acro{MAC}{media access control}
    \acro{TSN}{Time Sensitive Networking}
    \acro{NTP}{Network Time Protocol}
    \acro{COTS}{commercial off-the-shelf}
    \acro{PDV}{packet delay variation}
    \acro{LAN}{local area networks}
    \acro{SNMP}{Simple Network Management Protocol}
    \acro{SDN}{Software defined networking}
    \acro{NIC}{Network Interface Card}
    \acro{NC}{Network Calculus}
    \acro{MDR}{Maximum Drift Rate}
    \acro{TDD}{Time Division Duplex}
    \acro{NR}{New Radio}
    \acro{PHC}{PTP Hardware Clock}
    \acro{TT}{TimeTether}
    \acro{API}{Application Programming Interface}
    \acro{DNC}{Deterministic Network Calculus}
    \acro{SP}{strict-priority}
    \acro{IWRR}{Interleaved Weighted Round-Robin}
    \acro{RB}{resource blocks}
    \acro{SDR}{software-defined radio}
    \acro{FIR}{finite impulse response}
    \acro{TDL}{tapped delay line}
    \acro{USRP}{Universal Software Radio Peripheral}
    \acro{ADC}{analog to digital converter}
    \acro{RFNoC}{RF Network on Chip}
    \acro{FSPL}{free space path loss}
    \acro{FPGA}{Field Programmable Gate Array}
    \acro{DSP}{Digital Signal Processors}
    \acro{PDP}{Power-delay profile}
    \acro{V2X}{vehicle-to-anything}
    \acro{PER}{packet error ratio}
    \acro{RSSI}{received signal-strength indication}
    \acro{Mote}{Zolertia Re-Mote}
    \acro{SOTA}{state-of-the-art}
    \acro{PAPR}{peak-to-average power ratio}
    \acro{DAC}{digital to analog converter}
    \acro{MCS}{Modulation Coding Scheme}
    \acro{BLER}{Block Error Rate}
    \acro{RMS}{root mean square}
    \acro{BLER}{Block error rate}
    \acro{UE}{user equipment}
\end{acronym}

\title{OpenAirLink: Reproducible Wireless Channel Emulation using Software Defined Radios \\
\thanks{The authors acknowledge the financial support by the Bavarian State Ministry for Economic Affairs, Regional Development and Energy (StMWi) for the Lighthouse Initiative KI.FABRIK (Phase 1: Infrastructure as well as the research and development program under grant no. DIK0249). 
The authors also acknowledge the financial support by the Federal Ministry of Education and Research of Germany (BMBF) in the program of "Souverän. Digital. Vernetzt." joint project 6G-life, project identification number 16KISK002.}
\author{\IEEEauthorblockN{Yash Deshpande, Xianglong Wang, Wolfgang Kellerer\\
\IEEEauthorblockA{ Chair of Communication Networks, Technical University of Munich, Germany \\
Email: \{yash.deshpande, xianglong.wang@tum.de, wolfgang.kellerer\}@tum.de}}
}
}
\maketitle

\begin{abstract}
This paper presents OpenAirLink(OAL), an open-source channel emulator for reproducible testing of wireless scenarios.
OAL is implemented on off-the-shelf software-defined radios (SDR) and presents a smaller-scale alternative to expensive commercially available channel emulators. 
Path loss and propagation delay are the fundamental aspects of emulating a wireless channel. 
OAL provides a simple method to change these aspects in real-time. 
The emulator is implemented using a finite impulse response (FIR) filter. 
The FIR filter is written in Verilog and flashed on the SDRs Field Programmable Gate Array (FPGA). 
Most processing transpires on the FPGA, so OAL does not require high-performance computing hardware and SDRs. 
We validate the performance of OAL and demonstrate the utility of such a channel emulation tool using two examples. 
We believe that open-source channel emulators such as OAL can make reproducible wireless experiments accessible to many researchers in the scientific community. 
\end{abstract}


\acresetall

\section{Introduction}
\label{sec:Introdcution}

Experimental validation and evaluation of wireless technologies on hardware platforms are essential for their acceptance in the industry.  
While simulation and theoretical modeling can provide insights and predictions, they often make abstractions or assumptions that may not reflect the complexities of actual hardware. 
Thus, conducting experiments on hardware helps researchers to validate their assumptions, assess the accuracy of their predictions, or uncover discrepancies in their models.
Indeed, many wireless testbeds have been instrumental in advancing knowledge in the wireless domain \cite{testbed_blueprint, testbed_links_stability, powder_testbed, wireless_testbed, illinois_windtunnel_testbed, colosseum, aerpaw_testbed}.

The reproducibility of scientific experiments serves as a cornerstone for establishing the credibility and reliability of research findings. 
Experiments must be conducted in a controlled environment to be reproducible and for users to quickly identify the causation between a parameter and the output~\cite{dagstuhl_prof}.
This reproducibility is challenging in wireless systems. 
Many factors outside the experimenter's control may affect the quality of a wireless channel.
Multipath reflections or interference from unwanted sources affect the wireless signal and could degrade the reliability of the results \cite{testbed_links_stability, illinois_windtunnel_testbed}. 
Moreover, many lab-based setups can't conduct wireless mobility experiments spanning longer distances. 
Thus, the need for a channel emulator arises. 
These physical devices can be configured to distort, delay, and attenuate a wireless signal per the user's need. 
Instead of antennas, the user places the channel emulator between the transmitter and receiver and connects them with conductive cables. 

However, the cost and complexity of commercially available channel emulators make them inaccessible to institutions with modest means. 
They take time to set up and often require wireless networking expertise to operate. 
Researchers working at an application level such as robotics and control often lack this wireless networking expertise.  
OAL aims to bridge this gap. 
It aims to provide a fast-installation channel emulator where the hardware can later be repurposed once the experiment is done.  
On the other hand, OAL is less scalable and is meant to be used for smaller-scale scenarios. 
The industrial channel emulators are also "closed," meaning modifications require vendors and experts' support. 
Incorporating an open-source approach to the wireless channel emulator not only promotes transparency and accessibility but encourages widespread adoption and community-driven innovation \cite{opensource_review_ieee, morgan2013exploring_opensource}. 
Moreover, it helps improve reproducibility as the same emulator can be utilized when replicating experiments~\cite{bowman2023improving}. 

This paper presents OAL, an open-source wireless channel emulator, and its design and implementation. 
OAL is publicly available under the GNU License\footnote{https://github.com/n3martix/OpenAirLink}.
The paper also presents validation and testing of the performance of OAL and outlines the methodology to validate any such channel emulator.
The performance of OAL is compared with the Spirent Vertex, a \ac{SOTA} industrial channel emulator. 
Finally, examples of where OAL could be helpful, such as low-powered wireless networks based on IEEE 802.15.4 \cite{IEEE_standard} and in an Openairinterface \cite{oai_paper} 5G testbed, are demonstrated.


\section{Related Work}
\label{sec:related_work}

The most common approach for characterizing a wireless channel's path loss, delay, and delay spread is employing a \ac{TDL} model. 
The use of \ac{FIR} filters on \ac{DSP} and \ac{FPGA} to emulate the \ac{TDL} has been explored in many works \cite{lasko_old_delay, fpga_channel_emulator, eslamiDesign2009, chaudhari2018scalable}.
Practical channel models at the system level simulate or emulate, at the most, three paths \cite{channel_models_comparison, quadriga}.
Hence, clustering mechanisms approximate the \ac{PDP} of a large number of multipath coefficients to the few paths used for simulation/emulation \cite{clustering, rf_scenarios}. 
Thus, a sparse \ac{FIR} filter is more efficient in emulating a channel with a certain path delay and at most three multipath components~\cite{sparse_fir_filters, v2x_scenario, chaudhari2018scalable}.

Channel emulation on \ac{FPGA} and using \ac{FIR} filters has been proposed by \cite{first_emulator_repeatable, emulation_without_SDR}. 
However, these works are before the advent of \ac{SDR}s. 
Hence, these works had to develop their own RF frontend and connect it to an FPGA board. 
The \ac{SDR} hardware as shown in \figref{fig:emu-arch} integrates the RF frontend with digital signal processing and can retransmit the converted signal back through an output port~\cite{dillinger2005software}.
Thus, the \ac{SDR} helps build one channel emulator in a box. 

A large-scale channel emulator where \ac{SDR} is used as the RF frontend is proposed in \cite{chaudhari2018scalable}. 
However, this requires a complex \ac{FPGA} array in between the \ac{SDR} cards to deal with scale. 
This significantly pushes the cost of implementation up for such a system. 
Indeed, Colosseum \cite{colosseum} implements such a system at great cost and allows wireless researchers worldwide to connect to it. 
However, the idea behind OAL is for researchers to build and test smaller-scale wireless systems in their labs before testing them at scale in systems such as the Colosseum.
An emulator purely based on \ac{SDR}s to emulate a \ac{V2X} channel is presented in~\cite{v2x_scenario}. 
They also implement doppler effects and small-scale stochastic fading. 
However, the work evaluates only the \ac{PER} of an emulated system when using the IEEE 802.11p protocol. 
In this paper, we also discuss some fundamental specifications in such \ac{FPGA} based channel emulators, such as the dynamic range, resolution, and emulation precision, which are missing from the works mentioned above. 

Finally, for completeness, we mention other ways to deal with the repeatability of wireless channels. 
Random unwanted effects observed on the channel can be eliminated by detecting outliers after the experiment, as shown in~\cite{testbed_links_stability}.
The experiments can also be conducted in an anechoic chamber such as\cite{illinois_windtunnel_testbed}.
One could skip the RF-level channel emulation and abstract the effects to a higher level of emulation, such as IQ-level or packet-level emulation~\cite{aerpaw_testbed}. 
However, this restricts the type of experiments that can be conducted and the hardware that can be used in them~\cite{aerpaw_evaluation}.
Finally, the abstraction from RF to IQ or packet-level emulation also misses certain scenarios in the abstraction which could be useful to capture the full effect of the test.

\section{Implementation}
\label{sec:Implementation}

\subsection{Emulator Architecture} \label{sec:imp-arch}
\begin{figure}
    \centering
    \includegraphics[width=0.48\textwidth]{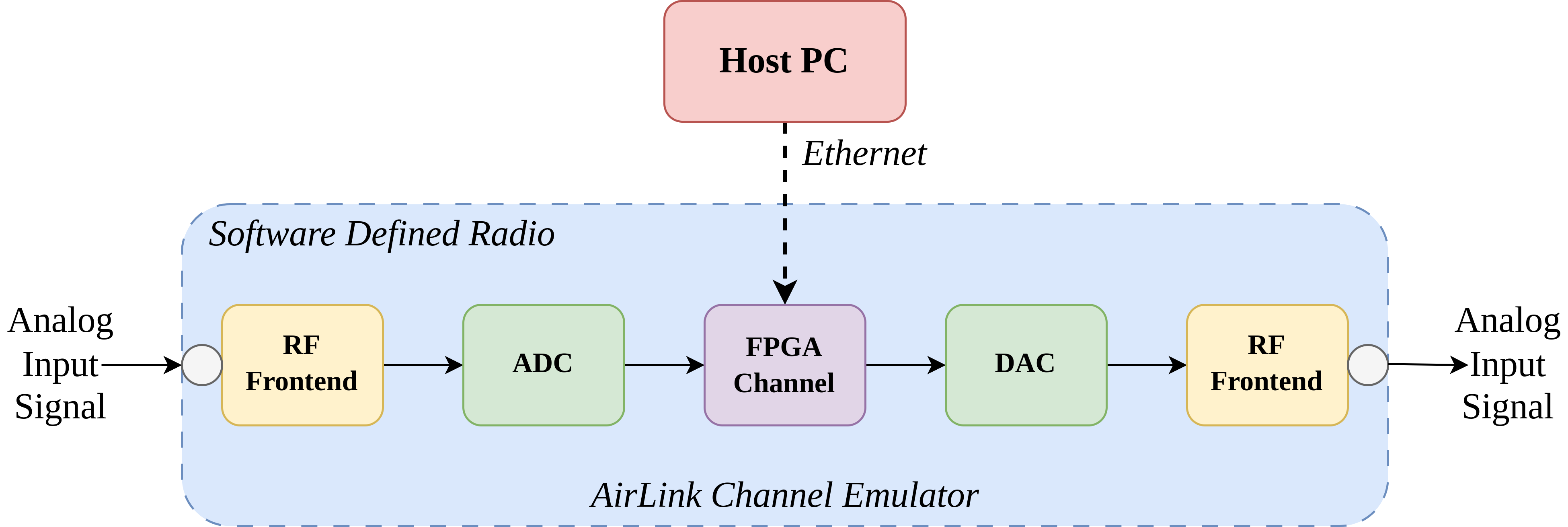}
    \caption{\small{Emulator Architecture: The software-defined radio (SDR) hardware is used to build the entire channel emulator, integrating the signal conversion hardware with an FPGA-based DSP. The host PC with an ethernet connection is used to control the delay and attenuation of the channel.} }
    \label{fig:emu-arch}
\end{figure}
\figref{fig:emu-arch} shows the overall architecture of the channel emulator. The RF frontend receives and downconverts the analog signal, which is then passed to the \ac{SDR}s \ac{ADC}.
The \ac{ADC} in the \ac{SDR} performs IQ sampling and quantization such that a continuous signal $X(t)$ is converted to a complex sample $\hat{X}[n] = \Delta \cdot \text{round} (\frac{X(nT_{s})}{\Delta})$, where $n\in\mathbb{N}$, $\hat{X}[n] \in \mathbb{C}$, $T_{s}$ is the sampling period and $\Delta$ is the quantization level. 
We get $\Delta$ by dividing the input signals range with the resolution of the \ac{ADC} $2^{R}$. 
The \ac{ADC}s sampling period $T_{s}$ and resolution $2^{R}$ provide a fundamental limit on the bandwidth and precision of the channel emulator. 
The \ac{ADC} then feeds its digital output to the emulated channel on the \ac{FPGA}. 
The channel is applied to the signal, and the digital samples output $\hat{Y}[n] \in \mathbb{C}$ are then re-converted to an analog signal $Y(t)$ by the \ac{DAC} and transmitted by the RF front end.

The \ac{SDR} connects to a host PC via Ethernet. 
The PC issues control commands and updates the emulated channel at run time. 
This architecture offers stable and low-latency pipeline processing by utilizing the \ac{SDR}'s \ac{FPGA} and performing all signal processing onboard.

\subsection{FPGA Channel}
The FPGA channel applies the desired channel fading, including delay and power loss for signal path and multi-path effect from reflection or diffraction to the RF signal. 
This requires an efficient model that works in FPGA to simulate effects. 
The theoretical model used for the emulator is a \ac{TDL} model. 
A \ac{TDL} model consists of delay-line and tap-output components. 
It simulates propagation delay and scales the signal correspondingly. Mixing single paths with different path delays can produce the multipath effect. 
It can precisely simulate various RF fading channels by designing delay and tap scale coefficients. 
The \ac{TDL} model is of order $N$ and produces the output

\begin{equation} \label{eq:tdl}
   \hat{Y}[n] = \sum_{i=0}^{N} b_{i}\cdot \hat{X}[n-i], 
\end{equation}

where $x[n-i]$ is the $i$-th \textit{tap} with a coefficient $b_{i}$ and $0\leq i \leq N$. 

A \ac{TDL} channel can be considered a general causal \ac{FIR} filter. 
The implementation of the emulator is powered by \ac{RFNoC}~\cite{rfnoc}, a framework used for implementing DSP in \ac{USRP}'s FPGA. 


\subsubsection{Path Delay}
Block shift registers advance the digital samples through the taps at each clock cycle. 
Therefore, the resolution of the path delay is defined by the FPGA clock cycle frequency, $f\cdot\si{Hz}$, giving a delay step size of $\frac{1}{f}\si{\s}$. 
With $N$ number of taps, such a channel emulator can support $\frac{N-1}{f}\si{\s}$ of maximum delay.
Considering that the RF signal propagates at the speed of light $c$ in free space, the step size of the path distance is $\frac{c}{f}\cdot\si{\m}$, with a total propagation distance of $\frac{c\cdot(N-1)}{f}\cdot\si{\m}$. 
The clock cycle also quantizes the distance between multipath components in the delay spread.
The minimum spread between any two components of the signal is limited to the delay step size of $\frac{1}{f}\si{\s}$. 

\subsubsection{Attenuation}
The IQ samples are scaled by the multiplier $b_{i} \in [-2^{r}, 2^{r}] \cap \mathbb{Z}$ inside the \ac{FIR} filter, where $r+1$ is the number of bits allocated for signed integers in the \ac{FPGA}.  
We are only interested in attenuating the signal. 
Therefore, the path gain for the $i$-th tap is, 

\begin{equation} \label{eq:att}
    G  = 20log_{10}(\frac{b_{i}}{2^{r}-1})
\end{equation}

where $b_{i} \in [1, 2^{r}-1] \cap \mathbb{N}$. 
From Eq.\ref{eq:att}, the resolution of attenuation is given by,

\begin{equation} \label{eq:att_step}
    \Delta G = 20log_{10}(\frac{b_{i}+1}{b_{i}}).
\end{equation}

Hence, attenuation's resolution degrades with the emulation distance between the transmitter and receiver.
We have approximately a dynamic range of attenuation from the two extreme values that $b_{i}$ can take, $D \approx r\cdot20log_{10}(2)$. This dynamic range of a few hundred meters (in free space) restricts the emulation capacity. 
Secondly, the resolution at the tail end of this dynamic range would be 6dB, meaning a distance resolution of a hundred meters in the case of the \ac{FSPL} model.
Hence, using a method similar to~\cite{fpga_channel_emulator}, we define the dynamic range given a maximum attenuation resolution $\Delta Gm$. 
First, we put $\Delta Gm$ in Eq.\ref{eq:att_step} to find the minimum value of $b_{i}$ that gives us the desired resolution. 
Then, we can find our reduced dynamic range for the given allowed resolution by subtracting the two extreme values that $b_{i}$ can take,
\begin{equation}\label{eq:reduced_dynamic_range}
    D_{\Delta Gm} = r\cdot20log_{10}(2) + 20log_{10}(10^{\frac{Gm}{20}}-1)   
\end{equation}

OAL employs a bit shift operation before the \ac{FIR} filter that coarsely attenuates the signal for all the path components. 
A right shift by $j$ bits is equivalent to dividing the sampled values by $2^j$ or attenuating the signal by approximately $6j$dB.  
If the maximum number of bit shifts is $s$, we get a higher dynamic range, 
\begin{equation}\label{eq:increased_dynamic_range}
    D_{\Delta Gm} = (r+s)\cdot20log_{10}(2) + 20log_{10}(10^{\frac{Gm}{20}}-1).  
\end{equation}

Thus, we get the worst-case resolution for our system by plugging $D_{\Delta Gm} = 6$dB in Eq. \ref{eq:reduced_dynamic_range} and solving for $Gm$. 
Putting it all together, we get attenuation resolution for when bit-shift is employed,  
\begin{equation}\label{eq:final_resolution}
    Gbs \approx 20log_{10}(\frac{10^{\frac{3}{10}}}{2^{r}} + 1).
\end{equation}

\begin{figure}
    \centering
    \includegraphics[width=0.99\columnwidth]{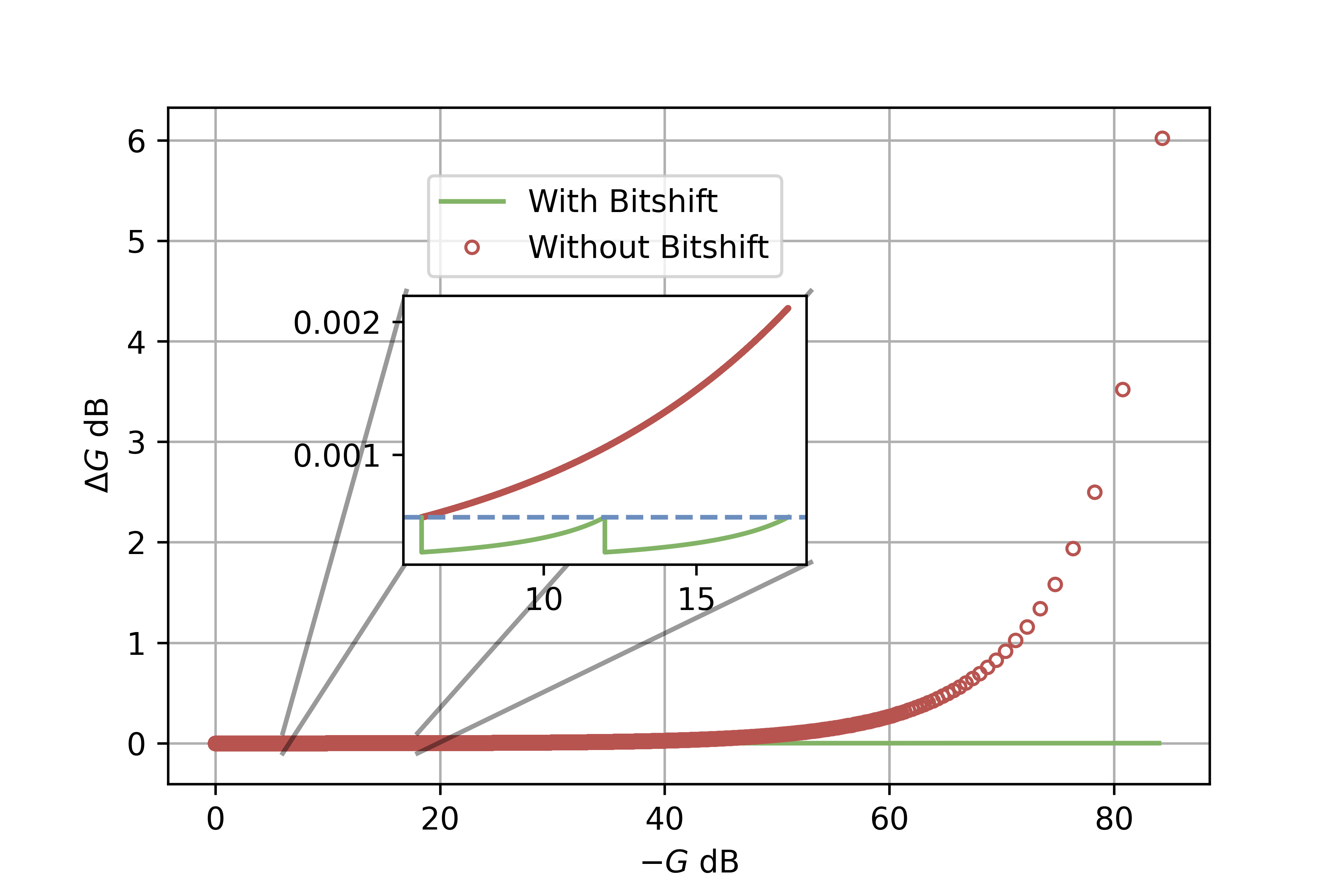}
    \caption{\small{Attenuation Resolution: As the attention increases, the resolution of the emulator degrades. For a $r=15$ bit division of the digital sample, a maximum attenuation of roughly 82 dB is possible. However, the resolution at this attenuation level is almost 6 dB. The bit-shifting operation can greatly improve this resolution. The worst-case resolution when bit shift is employed for the same value of $r$ is 0.000528 dB shown in the dashed blue line.}}
    \label{fig:chan-step}
\end{figure}

The effect of the bit-shift operation is shown in Fig.~\ref{fig:chan-step} for $r=15$.
The resolution from Eq~\ref{eq:att_step} is plotted against the desired attenuation $G$. 
The dynamic range without bit shifting is restricted to 80 dB, and the resolution worsens as we reach this value. 
The bit-shifting value increases the dynamic range well beyond 80 dB and the worst-case resolution $Gbs$ to 0.00058 dB. 
The zoomed portion of the plot shows how the resolution is bounded at $Gbs$ by the blue dashed line. 
Finally, we mention that the IQ output values in the \ac{FPGA} \ac{FIR} filters $\hat{Y}[n]$ can only be integers. 
Due to truncation, a small phase distortion of maximum value 0.5*LSB is introduced by the \ac{FPGA} channel for both the I and Q part of the sample. 






\subsection{Channel Update}
\label{subsec:channel_update}
Section ~\ref{sec:imp-arch} mentions that the emulator connects to the host PC through Ethernet. 
A program in the host PC controls the emulator and updates the $b_{i}$, where $i \in \{1,2,...,N\}$ coefficients on the \ac{FPGA} in runtime.
Similar to Coloseum \cite{colosseum}, the program that updates these coefficients reads a pre-prepared CSV file. 
Suppose the channel can be updated $u$ times per second, then the maximum velocity of the channel emulator at resolution $\frac{c}{f}\cdot\si{\m}$ is $\frac{c \cdot u}{f}\cdot\si{\m/\second}$. 
Thus, we can increase the maximum velocity of the nodes if we decrease the resolution. 


\subsection{Calibration}
\label{subsec:calibration}

The last step of implementation involves hardware calibration. 
For \ac{USRP}s, the RF frontend TX and RX gains can be adjusted via the UHD library tools.
The emulator is connected and set in a pass-through mode, i.e., no attenuation or delay is applied.
Then, a specific transmit power is set on the transmitter, and the received power must equal the sum of transmit power, the transmitter's TX gain, and the receiver's RX gain. 
If that's not the case, the RX gain of the RF frontend of OAL should be adjusted accordingly. 
Due to different values of the \ac{PAPR} and inherent losses in wireless hardware, this calibration must be done before every test setup. 
The \ac{USRP} also produces a tone with the oscillator frequency, which is caused by the DC offset due to the local oscillator power leakage when up/down-converting RF signals. 
A self-calibration program by the UHD library can minimize but not eliminate it. 

\section{Specifications}
\label{sec:specs}

The \ac{SDR} chosen for OAL is the \ac{USRP} X310.
This device is widely used and supported by many software frameworks, such as GNU Radio and MATLAB. 
The \ac{USRP} is equipped with a Kintex 7 FPGA that runs the emulated channel.
A signed integer on the Kintex 7 is 16 bits, giving the value of $r$ to be 15. 
The \ac{ADC} has a maximum sample rate $\frac{1}{T_{s}}$ of 200 MSample/s and a resolution $R$ of 14-bit/sample ADC, while the \ac{DAC} resolution is  16-bit/sample. 
The SBX-120 daughterboard is the RF front-end, with $120$MHz analog bandwidth and independent TX and RX modules, simultaneously emulating uplink and downlink channels by a single \ac{USRP}.
The daughterboard supports a frequency range of 400MHz to 4.4GHz. 
The 

\begin{table}
\centering
    \begingroup
    \renewcommand{\arraystretch}{1} 
    \begin{tabular}{|p{2.7cm}|p{1.5cm}|p{2.7cm}|}
        \hline
        \textbf{Parameter} & \textbf{Formula}  & \textbf{Value} \\ \hline
         Maximum Bandwidth & - & 120 MHz  \\ \hline
         Frequency Range & - & 400MHz to 4.4GHz \\ \hline
         Delay Resolution & $1/f\si{\s}$ & 5 $\si{\nano \s}$ \\ \hline
         Maximum Delay & $(N-1)/f\si{\s}$ &  205 $\si{\nano \s}$ \\ \hline
         Attenuation Resolution & Eq.\ref{eq:final_resolution} & 0.000528 dB \\ \hline
         Maximum Attenuation & Eq.\ref{eq:increased_dynamic_range} & $\approx$144 dB\\ \hline
         Maximum Velocity & $(c\cdot u)/f$ & 1500 $\si{\m/\s}$ \\ \hline
    \end{tabular}
    \endgroup
    \caption{\small{OAL Specifications: The achievable bandwidth, delay, attenuation resolution, and velocity.}}
    \label{tab:OpenAirlink_specs}
\end{table}
The specifications of the OAL are shown in Table\ref{tab:OpenAirlink_specs}. 
Theoretically, the bandwidth of the channel emulator depends on the Nyquist criterion based on the sampling rate of the \ac{ADC}. 
This value for the \ac{USRP} X310 is 200MHz since the sampling rate is 200 MSamples/s and the sampling method is IQ. 
However, OAL's bandwidth is limited by the bandwidth of the RF frontend SBX-120 daughterboard, which is 120 MHz.
The daughterboard also supports a frequency range of 400MHz - 4.4GHz. 
This value is large enough to cover most commercially known wireless systems, except for wideband systems. 

The \ac{FPGA} clock cycle frequency $f$ is 200 MHz. 
This gives us a delay resolution of 5 ns. 
In free space, that corresponds to a distance resolution of 1.5 m. 
The number of taps in the \ac{FIR} filter $N$ is 42, and hence we get a maximum delay of 205 ns or a maximum distance of emulation between transmitter and receiver in free space of 60 m. 
Further extending the number of taps $N$ and determining the limits of the filter size is a part of our future work. 
The attenuation resolution of 0.000528 dB and dynamic range of 144 dB are obtained from $r=15$ and $s=8$. 
These limits are placed due to the size of integers on the Kintex 7 and the truncation caused by integer values on the \ac{FPGA}.
Finally, we tested on a 4-core Intel i7 PC with 10G ethernet card, the effect of the channel update rate. 
At an update rate $u$ of 1000 updates per second, the maximum CPU utilization was just 54 percent. 
Thus we can control multiple OAL \ac{SDR}s from a single host PC. 
At the best distance resolution and update rate, the maximum velocity that AirLnk can support is 1500 m/s. 
This speed is already much more than what is needed for most mobility applications. 
It must be noted that OAL does not emulate doppler effects, which would play an essential role in higher velocity emulation. 
This is also a part of our future work.

\section{Verification and Performance Tests}
\label{sec:verification}

This section aims to verify OAL's operation, test its performance, and demonstrate that it can reliably reproduce a wireless channel.
The tests in this section utilize other \ac{USRP}s and inexpensive RF modules and can be reproduced for testing the performance of any such wireless channel emulator. 

\begin{figure}
    \centering
    \includegraphics[width=0.4\textwidth]{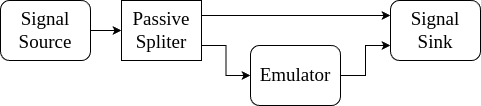}
    \caption{Verification Setup: A signal is split into two copies using a passive RF splitter. One copy goes directly to the signal sink and the other to the sink via the channel emulator. The comparison between the two copies allows us to test the latency of the OAL and verify the delay operation.}
    \label{fig:test-setup}
\end{figure}

The test setup is organized as shown in \figref{fig:test-setup}.
A test signal generated by the signal source is split into two identical copies using a passive RF splitter.
One copy is sent to the emulator for processing, while the other is used as a reference.
Both copies are recorded and compared at the signal sink.
The signal source and sink are the same \ac{SDR}s, namely the \ac{USRP} X310. 
All three \ac{SDR}s clocks are disciplined via an external PPS and 10MHz source using the NI Octoclock time distribution device to mitigate the effects of clock drifts in the latency measurements.

\subsection{Processing Latency and Delay}
\label{subsec:processing_latency}
The latency of the emulator refers to the time required for the RF signal to pass through the \ac{USRP} and apply channel effects.
The additional time OAL adds is measured by analyzing the difference between the test signal's time arrival from the two paths shown in \figref{fig:test-setup}.
The delay is derived from the negative of the lag at the point where the normalized cross-correlation between the two signals has the highest absolute value.
To determine each OAL component's latency, we first measured the processing latency through pass-through, where the received signal is immediately re-transmitted without applying the \ac{FPGA} channel.
We then enabled each component of our FPGA channel and measured the increased latency to demonstrate the extra time introduced by the channel components.
Around 52 percent of the latency is introduced by the RF frontend and ADC/DAC components, which have the tasks of signal trans-receive, up/down-conversion, and sampling.
The FPGA channel is implemented using RFNoC~\cite{rfnoc}, which constrains its processing latency performance due to the Module-NoCCore-Module structure with FIFO buffers in between to ensure pipeline processing and prevent sample loss due to buffer overflows.
 
 \begin{figure}
    \centering
    \includegraphics[width=0.4\textwidth]{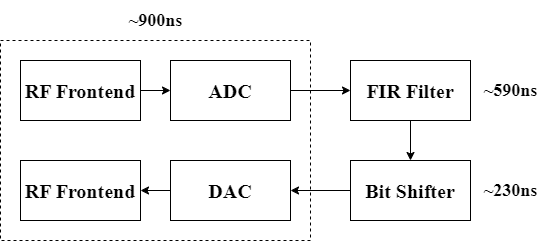}
    \caption{\small{OAL Processing Latency: The processing latency of OAL. Each component's variance in processing latency was maximum $\pm 1$ns. This certainty in the latency is helpful for the reproducibility of the wireless channel.}}
    \label{fig:emu-latency}
\end{figure}

\figref{fig:emu-latency} displays the measured processing latency, with a total time of approximately $1.72$us.
The measured latency demonstrated high stability with the maximum variance at each component bounded by $\pm1$ns.
We performed the same overall processing latency test for the Spirent Vertex channel emulator.
The results showed similar stability, but the magnitude of the processing latency was $\approx 3.19$us.
The datasheet of another \ac{SOTA} industrial channel emulator, namely the Keysight F8800A Propsim F64, states the value of processing latency to be $2.6$us. 
Hence, OAL stands out from other channel emulators regarding latency, although it lacks the cross-channel mixing usually available in industrial ones. 

Path delay refers to the emulator's capability to delay a signal to emulate wireless propagation delay.
We set the \ac{FIR} filter coefficients to emulate a given propagation delay and measure it using the same method used to measure the processing latency. 
As mentioned in Section~\ref{sec:specs}, the resolution for path delay for OAL is $5$ns.
We tested the path delay performance of OAL across the entire dynamic range of the emulator.
Once again, the results demonstrated the same stability as in the case of processing latency and perfect accuracy.




\subsection{Attenuation}
\label{subsec:attenuation}
The emulator's other core operation is to scale the RF signal with the desired attenuation.
We verify whether the received power at the signal sink matches the expected values after attenuation. 
At the same time, we conduct a conformance test to check if the signal is valid at reception after the emulation operation. 
Hence, instead of solely measuring and comparing the received signal power, we use IEEE 802.15.4 hardware transceivers as signal source and sink.
The attenuation test is conducted by \ac{Mote}\cite{zolertia} and Contiki OS \cite{Contiki}.
The \ac{Mote} carries a cc2538 system-on-chip microcontroller and runs an IEEE 802.15.4 radio with a build-in input \ac{RSSI} measurements.
From the data sheet \cite{cc2538}, the actual input power is given by $P = \text{RSSI} - \text{offset}$, where offset is the front-end gain set during system design and varies for microcontrollers, the value for the cc2538 is $73$dB.

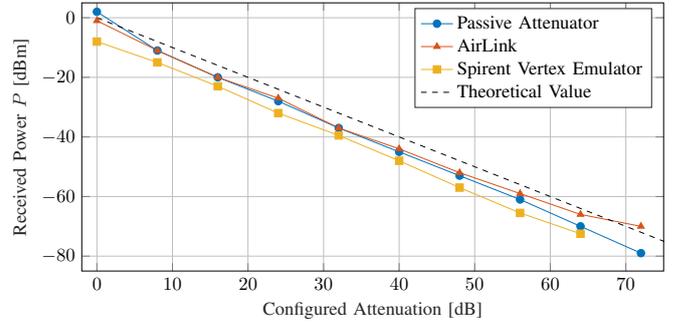
\begin{figure}
    \centering
    \scalebox{0.70}{
%
%
\definecolor{mycolor1}{rgb}{0.00000,0.44700,0.74100}%
\definecolor{mycolor2}{rgb}{0.85000,0.32500,0.09800}%
\definecolor{mycolor3}{rgb}{0.92900,0.69400,0.12500}%
\begin{tikzpicture}

\begin{axis}[%
width=4.306in,
height=2in,
at={(0.722in,0.418in)},
scale only axis,
unbounded coords=jump,
xmin=-2,
xmax=75, 
xlabel style={font=\color{white!15!black}},
xlabel={Configured Attenuation [dB]},
ymin=-85,
ymax=5,
ylabel style={font=\color{white!15!black}},
ylabel={Received Power $P$ [dBm]},
axis background/.style={fill=white},
xmajorgrids,
ymajorgrids,
legend style={legend cell align=left, align=left, draw=white!15!black}
]
\addplot [color=mycolor1, mark=*, mark options={solid, mycolor1}]
  table[row sep=crcr]{%
0	2\\
8	-11\\
16	-20\\
24	-28\\
32	-37\\
40	-45\\
48	-53\\
56	-61\\
64	-70\\
72	-79\\
80	-inf\\
};
\addlegendentry{Passive Attenuator}

\addplot [color=mycolor2, mark=triangle*, mark options={solid, mycolor2}]
  table[row sep=crcr]{%
0	-1\\
8	-11\\
16	-20\\
24	-27\\
32	-37\\
40	-44\\
48	-52\\
56	-59\\
64	-66\\
72	-70\\
80	-inf\\
};
\addlegendentry{AirLink}

\addplot [color=mycolor3, mark=square*, mark options={solid, mycolor3}]
  table[row sep=crcr]{%
0	-8\\
8	-15\\
16	-23\\
24	-32\\
32	-39.5\\
40	-48\\
48	-57\\
56	-65.5\\
64	-72.5\\
72	-inf\\
80  -inf\\
};
\addlegendentry{Spirent Vertex Emulator}

\addplot [color=black, dashed]
  table[row sep=crcr]{%
0	0\\
8	-8\\
16	-16\\
24	-24\\
32	-32\\
40  -40\\
48	-48\\
56	-56\\
64	-64\\
72	-72\\
80	-80\\
};
\addlegendentry{Theoretical Value}

\end{axis}
\end{tikzpicture}%
}
    \caption{\small{Attenuation Operation: The attenuation of OAL is compared with a passive variable attenuator and the Spirent Vertex. All the measured values are averages over 1000 measurements. The received power values are lower than the theoretical value due to unaccounted losses in the conducive cables.} }
    \label{fig:emu-att}
\end{figure}
 
\figref{fig:emu-att} compares OAL with a passive attenuator and the Spirent Vertex.
The figure plots the average received input power at the \ac{Mote} from the reported \ac{RSSI} over 1000 measurements at each configured attenuation level.
The maximum variance for all three attenuation methods was $\pm3$dB. 

All measured values of $P$ are lower than the expected theoretical value due to unaccounted losses in the conductor cables and contact.
The \ac{RMS} between the measured value and the expected theoretical value for OAL was $10.44$dBm, which is lower than the $15.17$dBm of the passive attenuation and $24.29$dBm for the Spirent Vertex.  

\subsection{Conformance Test with 5G NR}
\label{subsec:5gConformance}
\begin{figure}
    \centering
    \scalebox{0.65}{
%
%
\begin{tikzpicture}

\begin{axis}[%
width=4.844in,
height=2.039in,
at={(0.812in,0.43in)},
scale only axis,
unbounded coords=jump,
xmin=0.5,
xmax=11.5,
xtick={1,2,3,4,5,6,7,8,9,10,11},
xticklabels={63, 64, 65, 66, 67, 68, 69, 70, 71, 72, 73},
xlabel style={font=\color{white!15!black}},
xlabel={Path Loss [dB]},
ymin=-10,
ymax=110,
ylabel style={font=\color{white!15!black}},
ylabel={BLER [\%]},
axis background/.style={fill=white},
xmajorgrids,
ymajorgrids
]
\addplot [color=black, dashed, forget plot]
  table[row sep=crcr]{%
1	0\\
1	0\\
};
\addplot [color=black, dashed, forget plot]
  table[row sep=crcr]{%
2	0.0941\\
2	0.2202\\
};
\addplot [color=black, dashed, forget plot]
  table[row sep=crcr]{%
3	0.03575\\
3	0.0847\\
};
\addplot [color=black, dashed, forget plot]
  table[row sep=crcr]{%
4	0.2921\\
4	0.5751\\
};
\addplot [color=black, dashed, forget plot]
  table[row sep=crcr]{%
5	1.1012\\
5	1.8392\\
};
\addplot [color=black, dashed, forget plot]
  table[row sep=crcr]{%
6	10.6851\\
6	12.0719\\
};
\addplot [color=black, dashed, forget plot]
  table[row sep=crcr]{%
7	39.4407\\
7	42.1751\\
};
\addplot [color=black, dashed, forget plot]
  table[row sep=crcr]{%
8	65.497275\\
8	68.4769\\
};
\addplot [color=black, dashed, forget plot]
  table[row sep=crcr]{%
9	93.8511\\
9	95.3128\\
};
\addplot [color=black, dashed, forget plot]
  table[row sep=crcr]{%
10	99.9102\\
10	100.6569\\
};
\addplot [color=black, dashed, forget plot]
  table[row sep=crcr]{%
11	100.5897\\
11	100.9075\\
};
\addplot [color=black, dashed, forget plot]
  table[row sep=crcr]{%
1	0\\
1	0\\
};
\addplot [color=black, dashed, forget plot]
  table[row sep=crcr]{%
2	0.0002\\
2	0.0055\\
};
\addplot [color=black, dashed, forget plot]
  table[row sep=crcr]{%
3	0\\
3	0\\
};
\addplot [color=black, dashed, forget plot]
  table[row sep=crcr]{%
4	0.0037\\
4	0.03805\\
};
\addplot [color=black, dashed, forget plot]
  table[row sep=crcr]{%
5	0.2758\\
5	0.52485\\
};
\addplot [color=black, dashed, forget plot]
  table[row sep=crcr]{%
6	7.7871\\
6	8.7035\\
};
\addplot [color=black, dashed, forget plot]
  table[row sep=crcr]{%
7	35.9062\\
7	37.5488\\
};
\addplot [color=black, dashed, forget plot]
  table[row sep=crcr]{%
8	58.2206\\
8	61.3395\\
};
\addplot [color=black, dashed, forget plot]
  table[row sep=crcr]{%
9	88.3765\\
9	91.24745\\
};
\addplot [color=black, dashed, forget plot]
  table[row sep=crcr]{%
10	96.9965\\
10	98.4754\\
};
\addplot [color=black, dashed, forget plot]
  table[row sep=crcr]{%
11	100.1045\\
11	100.368175\\
};
\addplot [color=black, forget plot]
  table[row sep=crcr]{%
0.9375	0\\
1.0625	0\\
};
\addplot [color=black, forget plot]
  table[row sep=crcr]{%
1.9375	0.2202\\
2.0625	0.2202\\
};
\addplot [color=black, forget plot]
  table[row sep=crcr]{%
2.9375	0.0847\\
3.0625	0.0847\\
};
\addplot [color=black, forget plot]
  table[row sep=crcr]{%
3.9375	0.5751\\
4.0625	0.5751\\
};
\addplot [color=black, forget plot]
  table[row sep=crcr]{%
4.9375	1.8392\\
5.0625	1.8392\\
};
\addplot [color=black, forget plot]
  table[row sep=crcr]{%
5.9375	12.0719\\
6.0625	12.0719\\
};
\addplot [color=black, forget plot]
  table[row sep=crcr]{%
6.9375	42.1751\\
7.0625	42.1751\\
};
\addplot [color=black, forget plot]
  table[row sep=crcr]{%
7.9375	68.4769\\
8.0625	68.4769\\
};
\addplot [color=black, forget plot]
  table[row sep=crcr]{%
8.9375	95.3128\\
9.0625	95.3128\\
};
\addplot [color=black, forget plot]
  table[row sep=crcr]{%
9.9375	100.6569\\
10.0625	100.6569\\
};
\addplot [color=black, forget plot]
  table[row sep=crcr]{%
10.9375	100.9075\\
11.0625	100.9075\\
};
\addplot [color=black, forget plot]
  table[row sep=crcr]{%
0.9375	0\\
1.0625	0\\
};
\addplot [color=black, forget plot]
  table[row sep=crcr]{%
1.9375	0.0002\\
2.0625	0.0002\\
};
\addplot [color=black, forget plot]
  table[row sep=crcr]{%
2.9375	0\\
3.0625	0\\
};
\addplot [color=black, forget plot]
  table[row sep=crcr]{%
3.9375	0.0037\\
4.0625	0.0037\\
};
\addplot [color=black, forget plot]
  table[row sep=crcr]{%
4.9375	0.2758\\
5.0625	0.2758\\
};
\addplot [color=black, forget plot]
  table[row sep=crcr]{%
5.9375	7.7871\\
6.0625	7.7871\\
};
\addplot [color=black, forget plot]
  table[row sep=crcr]{%
6.9375	35.9062\\
7.0625	35.9062\\
};
\addplot [color=black, forget plot]
  table[row sep=crcr]{%
7.9375	58.2206\\
8.0625	58.2206\\
};
\addplot [color=black, forget plot]
  table[row sep=crcr]{%
8.9375	88.3765\\
9.0625	88.3765\\
};
\addplot [color=black, forget plot]
  table[row sep=crcr]{%
9.9375	96.9965\\
10.0625	96.9965\\
};
\addplot [color=black, forget plot]
  table[row sep=crcr]{%
10.9375	100.1045\\
11.0625	100.1045\\
};
\addplot [color=black, forget plot]
  table[row sep=crcr]{%
0.875	0\\
0.875	0\\
1.125	0\\
1.125	0\\
0.875	0\\
};
\addplot [color=black, forget plot]
  table[row sep=crcr]{%
1.875	0.0055\\
1.875	0.0941\\
2.125	0.0941\\
2.125	0.0055\\
1.875	0.0055\\
};
\addplot [color=black, forget plot]
  table[row sep=crcr]{%
2.875	0\\
2.875	0.03575\\
3.125	0.03575\\
3.125	0\\
2.875	0\\
};
\addplot [color=black, forget plot]
  table[row sep=crcr]{%
3.875	0.03805\\
3.875	0.2921\\
4.125	0.2921\\
4.125	0.03805\\
3.875	0.03805\\
};
\addplot [color=black, forget plot]
  table[row sep=crcr]{%
4.875	0.52485\\
4.875	1.1012\\
5.125	1.1012\\
5.125	0.52485\\
4.875	0.52485\\
};
\addplot [color=black, forget plot]
  table[row sep=crcr]{%
5.875	8.7035\\
5.875	10.6851\\
6.125	10.6851\\
6.125	8.7035\\
5.875	8.7035\\
};
\addplot [color=black, forget plot]
  table[row sep=crcr]{%
6.875	37.5488\\
6.875	39.4407\\
7.125	39.4407\\
7.125	37.5488\\
6.875	37.5488\\
};
\addplot [color=black, forget plot]
  table[row sep=crcr]{%
7.875	61.3395\\
7.875	65.497275\\
8.125	65.497275\\
8.125	61.3395\\
7.875	61.3395\\
};
\addplot [color=black, forget plot]
  table[row sep=crcr]{%
8.875	91.24745\\
8.875	93.8511\\
9.125	93.8511\\
9.125	91.24745\\
8.875	91.24745\\
};
\addplot [color=black, forget plot]
  table[row sep=crcr]{%
9.875	98.4754\\
9.875	99.9102\\
10.125	99.9102\\
10.125	98.4754\\
9.875	98.4754\\
};
\addplot [color=black, forget plot]
  table[row sep=crcr]{%
10.875	100.368175\\
10.875	100.5897\\
11.125	100.5897\\
11.125	100.368175\\
10.875	100.368175\\
};
\addplot [color=black, forget plot]
  table[row sep=crcr]{%
0.875	0\\
1.125	0\\
};
\addplot [color=black, forget plot]
  table[row sep=crcr]{%
1.875	0.0239\\
2.125	0.0239\\
};
\addplot [color=black, forget plot]
  table[row sep=crcr]{%
2.875	0.0004\\
3.125	0.0004\\
};
\addplot [color=black, forget plot]
  table[row sep=crcr]{%
3.875	0.11335\\
4.125	0.11335\\
};
\addplot [color=black, forget plot]
  table[row sep=crcr]{%
4.875	0.8749\\
5.125	0.8749\\
};
\addplot [color=black, forget plot]
  table[row sep=crcr]{%
5.875	9.7551\\
6.125	9.7551\\
};
\addplot [color=black, forget plot]
  table[row sep=crcr]{%
6.875	38.78875\\
7.125	38.78875\\
};
\addplot [color=black, forget plot]
  table[row sep=crcr]{%
7.875	63.6763\\
8.125	63.6763\\
};
\addplot [color=black, forget plot]
  table[row sep=crcr]{%
8.875	92.7812\\
9.125	92.7812\\
};
\addplot [color=black, forget plot]
  table[row sep=crcr]{%
9.875	99.4145\\
10.125	99.4145\\
};
\addplot [color=black, forget plot]
  table[row sep=crcr]{%
10.875	100.5897\\
11.125	100.5897\\
};
\addplot [color=black, only marks, mark=asterisk, mark options={solid, draw=blue}, forget plot]
  table[row sep=crcr]{%
nan	nan\\
};
\addplot [color=black, only marks, mark=asterisk, mark options={solid, draw=blue}, forget plot]
  table[row sep=crcr]{%
2	0.2431\\
2	0.2753\\
2	0.3334\\
2	0.3476\\
2	0.518\\
};
\addplot [color=black, only marks, mark=asterisk, mark options={solid, draw=blue}, forget plot]
  table[row sep=crcr]{%
3	0.1162\\
3	0.2188\\
3	0.3001\\
3	0.3333\\
3	0.3333\\
};
\addplot [color=black, only marks, mark=asterisk, mark options={solid, draw=blue}, forget plot]
  table[row sep=crcr]{%
nan	nan\\
};
\addplot [color=black, only marks, mark=asterisk, mark options={solid, draw=blue}, forget plot]
  table[row sep=crcr]{%
5	2.1468\\
5	2.2358\\
};
\addplot [color=black, only marks, mark=asterisk, mark options={solid, draw=blue}, forget plot]
  table[row sep=crcr]{%
6	2.1742\\
6	4.5493\\
};
\addplot [color=black, only marks, mark=asterisk, mark options={solid, draw=blue}, forget plot]
  table[row sep=crcr]{%
7	16.599\\
7	23.0259\\
7	31.6647\\
7	32.9565\\
7	34.2736\\
7	42.8596\\
};
\addplot [color=black, only marks, mark=asterisk, mark options={solid, draw=blue}, forget plot]
  table[row sep=crcr]{%
8	40.6387\\
8	53.7612\\
8	54.9727\\
};
\addplot [color=black, only marks, mark=asterisk, mark options={solid, draw=blue}, forget plot]
  table[row sep=crcr]{%
9	66.0395\\
9	68.259\\
9	87.0299\\
};
\addplot [color=black, only marks, mark=asterisk, mark options={solid, draw=blue}, forget plot]
  table[row sep=crcr]{%
10	93.115\\
10	95.801\\
};
\addplot [color=black, only marks, mark=asterisk, mark options={solid, draw=blue}, forget plot]
  table[row sep=crcr]{%
11	98.876\\
11	98.9363\\
11	99.0048\\
11	99.2454\\
11	99.6023\\
11	99.6801\\
};
\end{axis}
\end{tikzpicture}%
}
    \caption{\small{5G NR Signal Conformance: The Block Error Rate increases with higher attenuation, culminating in a link failure.
    The MCS was kept constant throughout the measurements to 6. The NR gNB reports the BLER every second, and 200 such measurements were conducted for each path loss setting.} }
    \label{fig:emu-con}
\end{figure}
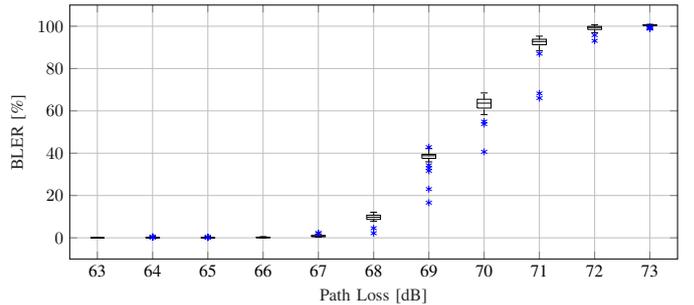

The attenuation verification conducted in Section \ref{subsec:attenuation} also demonstrates that OAL can be used for IEEE 802.15.4 signals.
OAL's design objective is to accommodate a spectrum of RF signals and protocols.
To demonstrate the versatility of OAL, we measured the uplink \ac{BLER} in an OpenAirInterface\cite{oai_paper} testbed.
A \ac{USRP} B210 is employed as the gNB, while a \ac{USRP} B210 mini is used at the \ac{UE}.  
In 5G, the devices adapt their \ac{MCS} depending on the channel quality, which leads to the \ac{BLER} fluctuating due to this adaptation. 
Hence, we fix the \ac{MCS} to 6 for this test and apply different attenuation values to the uplink signal. 
OpenAirInterface gNB reports the \ac{BLER} every second, and we measure 200 such measurement reports for one test.
\figref{fig:emu-con} shows an increase in the \ac{BLER} concurrent with increments in path loss.
These results substantiate OAL's adeptness at simulating 5G NR signal propagation, thus endorsing its utility in a broad spectrum of applications.

\section{Conclusion}
\label{sec:conclusion}

This paper presents OAL an open-source channel emulator realized with \ac{SDR}s.
Controlling the wireless channel quality in various experiments is crucial for the reproducibility of scientific work. 
Thus, wireless channel emulators are used to have a controlled RF environment and to emulate larger distances in laboratory rooms. 
OAL aims to provide a small-scale channel emulator to emulate point-to-point wireless links. 
The paper presents the \ac{FIR} implementation of OAL on the \ac{SDR}s \ac{FPGA} and OAL's specifications. 
The test and validation results show that OALs performance is comparable to or better than \ac{SOTA} industrial wireless channel emulators. 
The variance in emulating path delay and attenuation with OAL is minimal, demonstrating its utility in reliably reproducing a wireless channel. 
The paper demonstrates that OAL can preserve the symbol quality from various wireless protocols, proving its versatility.  
To the best of our knowledge, we find that OAL offers at least a 10x cost reduction while providing the same quality and reproducibility of wireless channels as compared to \ac{SOTA} industrial channel emulators.

\bibliographystyle{IEEEtran}
\bibliography{references,internal_references}

\end{document}